\documentclass[10pt,twoside]{classe-cmb}
\usepackage{amssymb,amsbsy,amsmath,amsfonts,amssymb,amscd}
\usepackage{latexsym,euscript,exscale,epic,eepic,epsfig}
\usepackage[english,francais]{babel}
\usepackage{times}
\def\eg{{\it e.g. }}
\def\etal{{et~al.}}
\def\ie{{\it i.e. }}

\def\zu{\rm\,}     

\def\microKCMB{$\mu\mathrm K_{CMB}$}

\newcommand{\OmL}{\Omega_{\Lambda}}

\newcommand{\OmT}{\Omega_{\rm tot}}

\newcommand{\Ombhh}{\Omega_{\rm b}h^2}

\newcommand{\Archeops}{\mbox{\sc{Archeops }}}
\newcommand{\Boomerang}{\mbox{\sc{Boomerang }}}
\newcommand{\Cobe}{\mbox{\sc{Cobe }}}
\newcommand{\Cbi}{\mbox{\sc{Cbi }}}
\newcommand{\Dasi}{\mbox{\sc{Dasi }}}
\newcommand{\Dmr}{\mbox{\sc{Dmr }}}
\newcommand{\Hfi}{\mbox{\sc{Hfi }}}

\newcommand{\Maxima}{\mbox{\sc{Maxima }}}
\newcommand{\Planck}{\mbox{\sc{Planck }}}

\newcommand{\Vsa}{\mbox{\sc{Vsa }}}
\def\ApJ{{\sl ApJ.}}
\def\ApJL{{\sl ApJ. Lett.}}
\def\AA{{\sl A\&A.}}

\def\MNRAS{{\sl MNRAS}}

\newtheorem{e-proposition}[theorem]{Proposition}

\newtheorem{e-definition}[theorem]{Definition\rm}

\ComParit{S1296-2147}
\PIT{FLA}
\PXHY{????}
\Add{?}
\Volume{0}
\Year{2003}
\FirstPage{1}
\LastPage{??}
\AuteurCourant{J.-Ch. Hamilton, A.~Beno\^{\i}t and the Archeops Collaboration}
\TitreCourant{Archeops results} 
\Journal
\Rubrique{Rubrique}{Heading}
\SousRubrique{Sous-rubrique}{Sub-Heading}
\PresentePar{Jean-Christophe HAMILTON, Alain BENOIT \& the Archeops Collaboration}{}
\Recu{}{}{}
\begin{document}
\selectlanguage{english}
\TitleOfDossier{The Cosmic Microwave Background}
\title{%
Archeops results
}
\author{%
Jean-Christophe HAMILTON~$^{\text{a,b,c}}$, Alain BENOIT~$^{\text{d}}$  and the Archeops collaboration~$^{\text{e}}$
}
\address{%
\begin{itemize}\labelsep=2mm\leftskip=-5mm
\item[$^{\text{a}}$]
LPNHE (CNRS-IN2P3), Paris VI \& VII - 4, Place Jussieu, 75252 Paris Cedex 05 - France\\
E-mail: hamilton@in2p3.fr
\item[$^{\text{b}}$]
LPSC (CNRS-IN2P3), 53, Avenue des Martyrs, 38026 Grenoble Cedex - France\\
\item[$^{\text{c}}$]
PCC (CNRS-IN2P3), 11, Place Marcelon Berthelot, 75231 Paris Cedex 05 - France
\item[$^{\text{d}}$]
CRTBT (CNRS),  25 Avenue des Martyrs, BP 166, 38042 GRENOBLE cédex 9 - France\\
E-mail: benoit@grenoble.cnrs.fr
\item[$^{\text{e}}$]
{\tt http://www.archeops.org}
\end{itemize}
}
\maketitle
\thispagestyle{empty}
\begin{Abstract}{%
  \Archeops is a balloon--borne instrument dedicated to measuring
  cosmic microwave background (CMB) temperature anisotropies at high
  angular resolution ($\sim$ 12 arcmin.) over a large fraction (30\%)
  of the sky in the (sub)millimetre domain (from 143 to 545~GHz). We
  describe the results obtained during the last flight: the \Archeops
  estimate of the CMB angular power spectrum linking for the first
  time \Cobe scales and the first acoustic peak, consequences in terms
  of cosmological parementers favouring a flat-$\Lambda$ Universe. We
  also present the first measurement of galactic dust polarization and
  accurate maps of the galactic plane diffuse (sub) millimetre
  emisson.}
\end{Abstract}

\par\medskip\centerline{\rule{2cm}{0.2mm}}\medskip
\setcounter{section}{0}
\selectlanguage{english}
\section*{Introduction}
The Cosmic Microwave Background (CMB) gives many clues as to the
origin of the Universe. It contains a wealth of diverse information,
in contrast with the other 2 so--called pillars of Cosmology. The
advent of BLIP (Background limited Performance) detectors (bolometers
at 100 to 300~mK) and mostly sidelobe--free HEMT based interferometers
has provided CMB maps with increasing accuracy and resolution in the
last 10 years. The fluctuations that are now routinely detected in a
few hours--days of integration time (\eg
\cite{boom1,boom2,maxima1,maxima2,dasi,cbi,vsa}) provide vivid
proof of the seeds that lead to large--scale structure formation. They
are best analysed with spherical harmonic angular power spectrum
$C_\ell$ as a function of multipole $\ell$ familiar to quantum
physics. The generation of the power spectrum is now theoretically
understood so that cosmological parameters can be deduced accurately.
\Archeops\footnote{{\tt http://www.archeops.org}} is a CMB
bolometer--based instrument with \Planck --
\Hfi\footnote{{\tt http://astro.estec.esa.nl/Planck}} technology that
fills a niche where previous experiments were unable to provide strong
constraints. Namely, \Archeops seek to join the gap in $\ell$ between
the large angular scales as measured by \Cobe/\Dmr and degree--scale
experiments, typically for $\ell$ between 10 and 200.  For that
purpose, a large sky coverage is needed. The solution was to adopt a
spinning payload mostly above the atmosphere, scanning the sky in
circles with an elevation of around 41~degrees.  The Earth's rotation
makes the circle span a large area of the sky.

In section~\ref{instru} we will describe the instrument
characteristics and the various flights, in
section~\ref{data_analysis} the data analysis process. We will present
the results concerning the CMB angular power spectrum in
section~\ref{cl} and the cosmological constraints in
section~\ref{cosmo}. The \Archeops diffuse galactic emission maps is
presented in section~~\ref{maps} and the first detection of galactic
dust polarization will be presented in section~\ref{polar}

\section{Instrument and flights\label{instru}}
The instrument~\cite{trapani} was designed by
adapting concepts put forward for \Planck -- \Hfi and using
balloon--borne constraints: namely, an open $^3$He--$^4$He dilution
cryostat cooling spiderweb--type bolometers at 100~mK, cold individual
optics with horns at the different temperature stages (0.1, 1.6, 10~K)
and the telescope. The Gregorian off--axis aluminum telescope is made of
an effective 1.5~m aperture primary and a secondary ellipsoid mirror. The
whole instrument is baffled so as to avoid stray radiation from the
Earth and the balloon.  The scan strategy imposes to observe by night.
Maximising integration time means going above the Arctic circle. After
a test flight in Trapani (Sicily) with four--hours integration time,
the upgraded instrument was launched three times from the Esrange base
near Kiruna (Sweden) by the CNES in the last 2 Winter seasons. The
last and best flight on Feb.~7th, 2002 yields 12.5~hours of CMB--type
data (at ceiling altitude and by night) from a 19--hours total. The
balloon landed in Siberia and it was recovered (with its precious data
recorded on--board) by a Franco--Russian team with --40~deg.C.
weather.

\section{Data analysis\label{data_analysis}}
The data are calibrated with the CMB dipole~\cite{smoot}, the FI\-RAS
Galaxy and Jupiter (point--like) emission (beside yielding effective
beams of typically 12~arcminute FWHM). The calibration error on the
dipole is estimated as 4\% (resp.~8\%) at 143 (resp.  217~GHz). The
other methods have calibration uncertainties of about 10~\% and are
consistent with the dipole calibration within less than 20~\%. Eight
detectors at 143 and 217~GHz are found to have a sensitivity better
than 200~\microKCMB\ rms, in one second of integration corresponding to
the stationary part of the noise. For a square pixel of 20~arcmin the
average 1~$\sigma$ sensitivity with all detectors combined per channel
is 100 and 150
\microKCMB\, (0.04 and 0.06 MJy/sr) at resp. 143 and 217~GHz.  It is
0.4 and 0.8~MJy/sr at 353 and 545~GHz. A large part of the data
reduction was devoted to removing additional noises which come from
the various thermal stages at frequencies $f\le 0.03 \zu Hz$, and
atmospheric effects: an elevation systematic effect is seen below
$0.1\zu Hz$ and the four frequencies are correlated between 0.1 and
1~Hz. These decorrelations in the timelines are mostly done for the
low $\ell<30$ side of the power spectrum.
The data are cleaned and calibrated, and the pointing is
reconstructed from stellar sensor data~\cite{benoit_pipe}. 

\begin{figure}[h!]
\resizebox{\hsize}{!}
{\includegraphics[clip,angle=90]{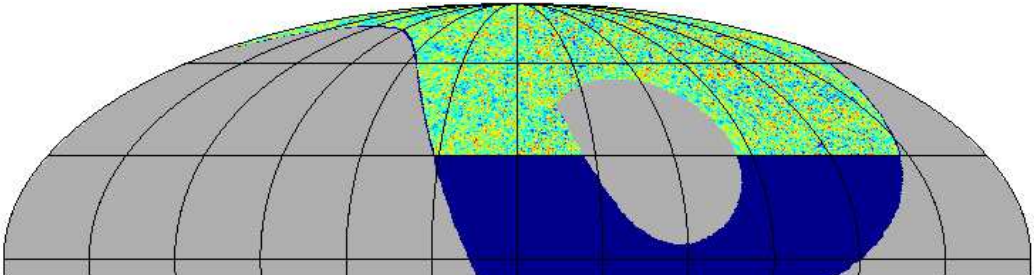} }
\caption{Archeops CMB map (Galactic coordinates, north hemisphere) in HEALPIX
  pixelisation~(\cite{healpix}) with 13.7~arcminutes pixels and a 15
  arcminutes Gaussian smoothing. The map was made by coadding the data
  from 2 photometers as discussed in the text.  The dark blue region
  is not included in the present analysis because of possible
  contamination by dust.  The Galactic anticenter is at the center of
  the map. The colors in the map range from -500 to 500
  $\mu\mathrm{K_{CMB}}$.\label{figure_map}}
\end{figure}

\begin{figure}[!b]
\resizebox{\hsize}{!}
{
\includegraphics[clip]{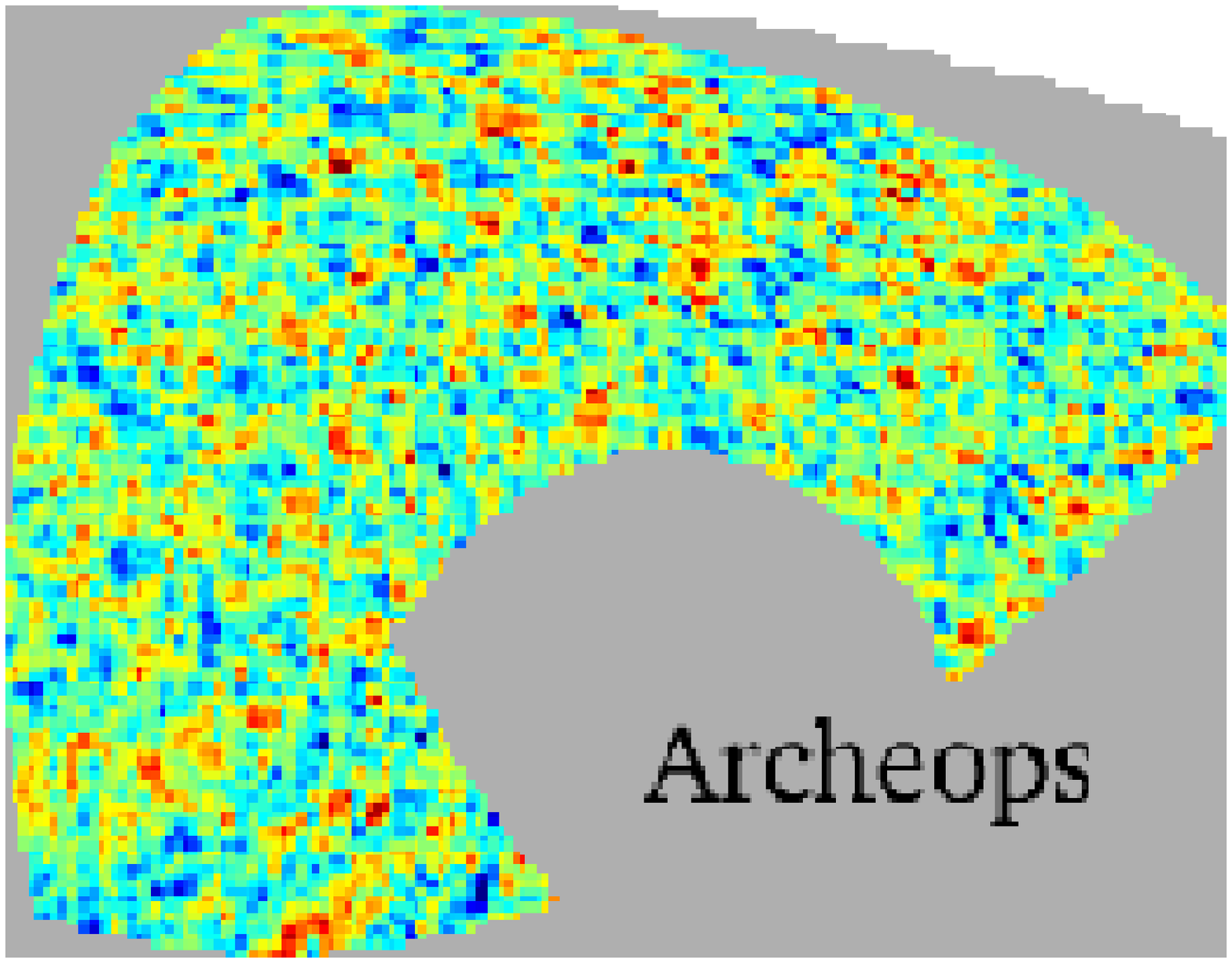}
\includegraphics[clip]{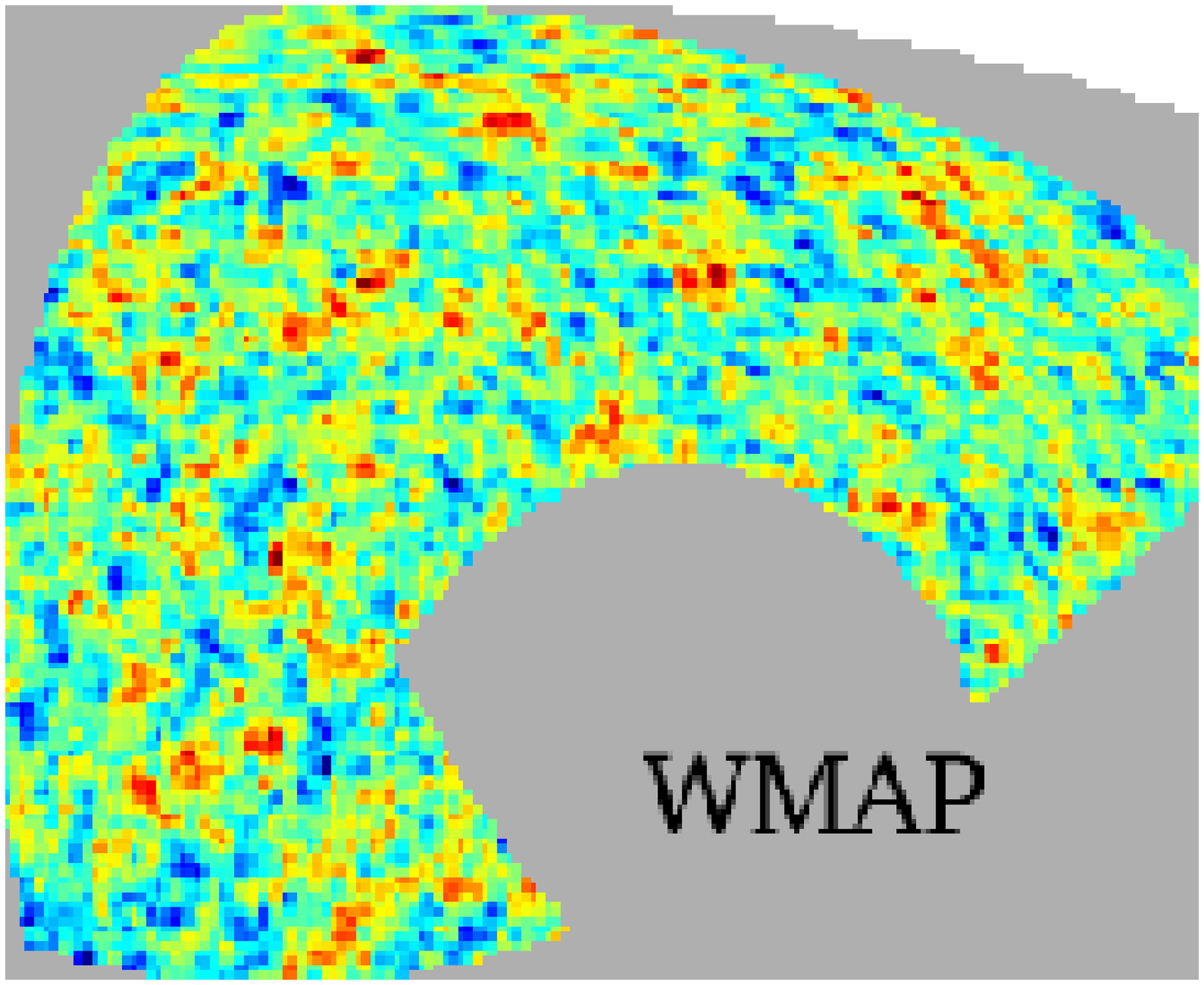} 
\includegraphics[clip]{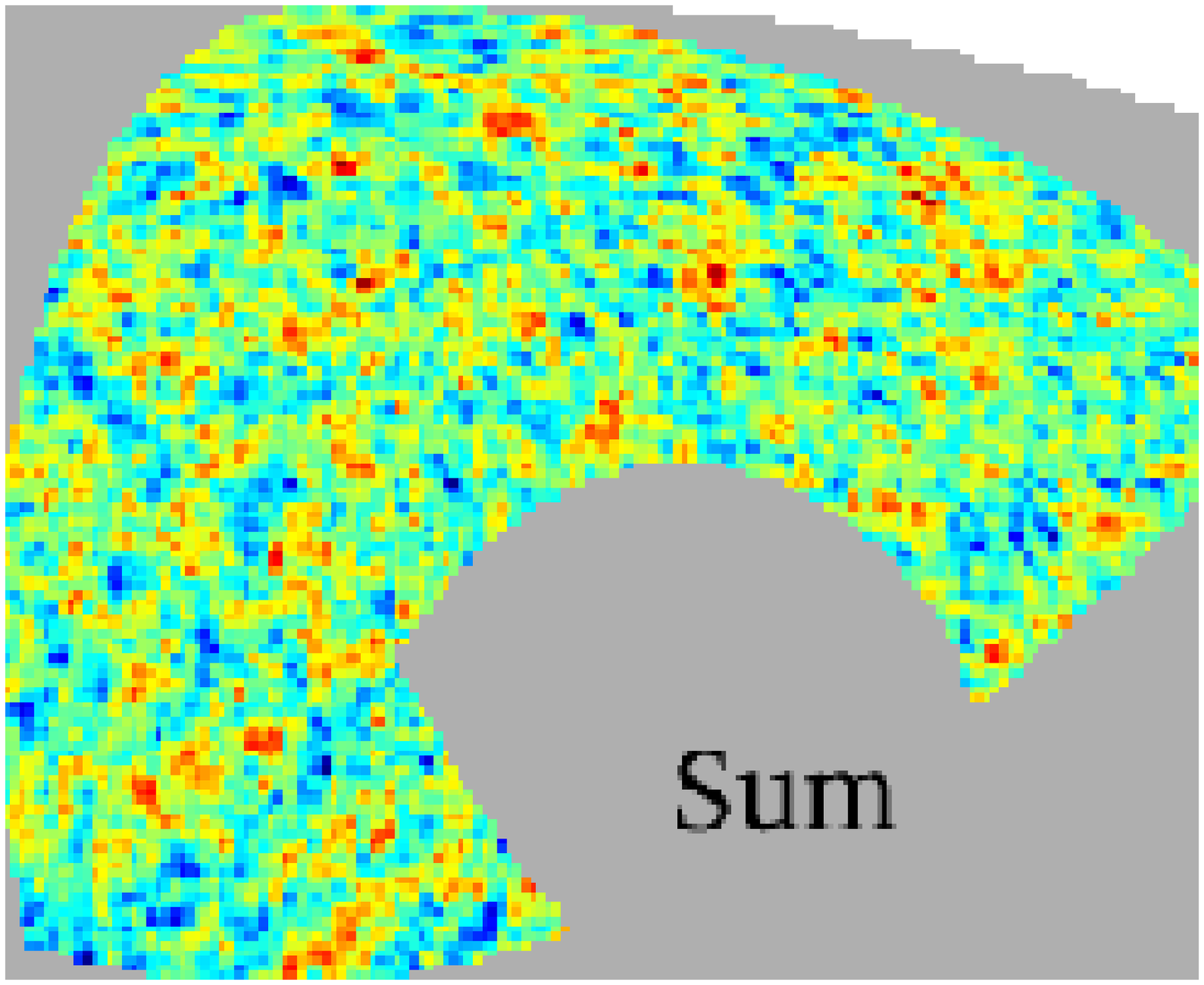}
\includegraphics[clip]{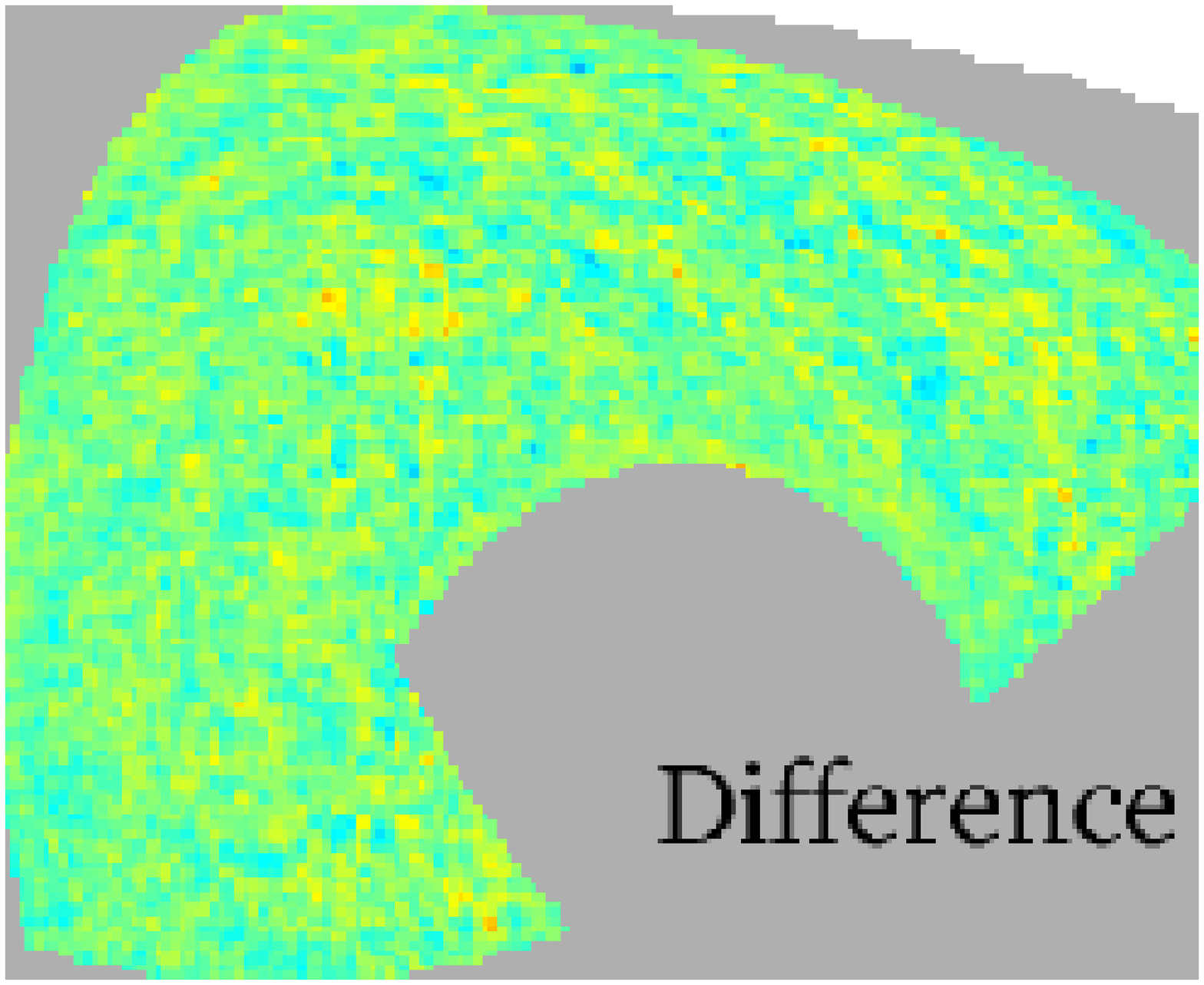} 
}
\centerline{\resizebox{6cm}{!}{\includegraphics[clip]{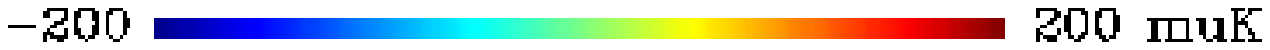}}}
\caption{Comparison of the Archeops and WMAP datasets. The first map
is from Archeops (convolved with a Gaussian beam of FWHM=1~deg. and 
rotated to help visualisation).
The second map is the same part of the sky from WMAP~\cite{wmap_map}. It is also convolved to a 1~deg. 
FWHM and large scales not present in the Archeops map are removed by 
subtracting the WMAP map convolved to a 15 deg.~FWHM. 
map. The third and fourth maps show the half sum and  half 
difference, both are performed without any intercalibration. 
The strong correlation between both datasets is obvious.
\label{compare_wmap}}
\end{figure}

We then make CMB-designed maps in the cleanest region (from the
Galactic foreground point of view) restricted to $b>+30$deg. giving a
total of $\sim 10^{5}$ 13.7 arcmin. pixels (HEALPIX nside=256) covering
12.6\% of the sky (see Fig.~\ref{figure_map}). The maps are obtained
by coadding filtered data (bandpass between 0.3 and 45 Hz) avoiding
ringing effects with a process described in~\cite{amblard}. Residual
ringing is estimated to be less than $\sim 36\mu\mathrm{K}^2$ on the
power spectrum in the first $\ell$-bin and negligible for larger
multipole. We then only keep maps of two best photometers, one at
143GHz and one at 217GHz which are coadded using a $1/\sigma^2$
weighting ($\sigma$ being the noise RMS in each pixel).  We apply the
MASTER method~\cite{hivon} to estimate the CMB power spectrum from
these maps. The noise Fourier power spectrum is estimated from the
time stream data using the algorithm described in~\cite{amblard}. Our maps
show structure on the sky at the degree angular scale that can be attributed to CMB anisotropies. A comparison of the anisotropies seen by Archeops and by WMAP is shown in Fig.~\ref{compare_wmap}.

\section{Angular power spectrum results\label{cl}}
\begin{figure}[h!]
\begin{minipage}{6cm}
\centering
\begin{tabular}{ccc@{~$\pm$~}c}\hline\hline
$\ell_{min}$ &$\ell_{max}$
&\multicolumn{2}{c}{$\frac{\ell(\ell+1)C_\ell}{(2\pi)}~(\mu\mathrm{K})^2$} \\
\hline
 15 & 22 &$  789$ &$  537$ \\
 22 & 35 &$  936$ &$  230$ \\
 35 & 45 &$ 1198$ &$  262$ \\
 45 & 60 &$  912$ &$  224$ \\
 60 & 80 &$ 1596$ &$  224$ \\
 80 & 95 &$ 1954$ &$  280$ \\
 95 &110 &$ 2625$ &$  325$ \\
110 &125 &$ 2681$ &$  364$ \\
125 &145 &$ 3454$ &$  358$ \\
145 &165 &$ 3681$ &$  396$ \\
165 &185 &$ 4586$ &$  462$ \\
185 &210 &$ 4801$ &$  469$ \\
210 &240 &$ 4559$ &$  467$ \\
240 &275 &$ 5049$ &$  488$ \\
275 &310 &$ 3307$ &$  560$ \\
310 &350 &$ 2629$ &$  471$ \\
\hline
\end{tabular}
\end{minipage}
\begin{minipage}{7.2cm}
\centering
\resizebox{8.6cm}{!}
{\includegraphics[width=5cm]{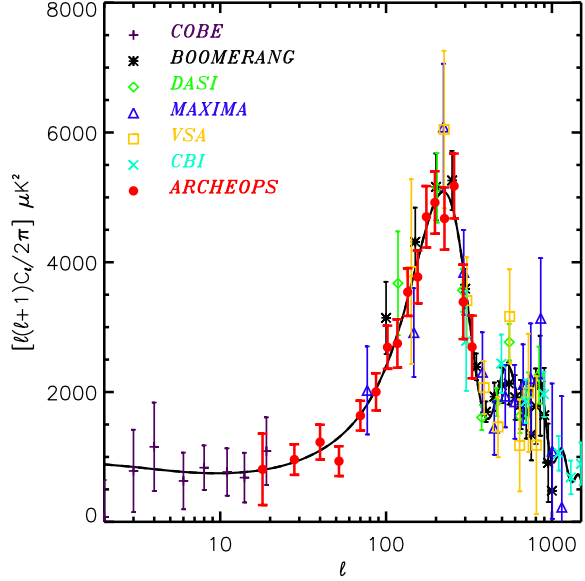} }
\end{minipage}
\caption{The Archeops CMB power spectrum for the best two photometers
(third column). The error bars are given at $1\sigma$. Data points
given in this table correspond to the red points in the figure.
\label{figure_cl}}
\end{figure}

 The CMB anglar power spectrum is estimated in 16 bins ranging from
 $\ell=15$ to $\ell=350$ (see Fig.~\ref{figure_cl} for a comparison
 with a selection of other recent experiments and a best--fit
 theoretical model). Much attention was paid to the possible
 systematic effects that could affect the results.  At low $\ell$,
 dust contamination and at large $\ell$, bolometer time constant and
 beam uncertainties are all found to be negligible with respect to
 statistical error bars. The sample variance at low $\ell$ and the
 photon noise at high $\ell$ are found to be a large fraction of the
 final \Archeops error bars in Fig.~\ref{figure_cl}. Various tests for
 systematic effects have been performed, such as Jack-knife tests and
 are described in~\cite{benoit_cl}.

The first acoustic peak appears clearly from \Archeops data
itself. A Gaussian fit of the Archeops and the COBE data leads to a 
location of the peak at $\ell_\mathrm{peak}=220\pm 6$ with a width 
$FWHM=192\pm 12$ and an amplitude $\delta T=71.5\pm 2.0\mu$K, compatible 
with previous determinations and achieving
a better precision than earlier experiments, even combined together.
One of the main goals of the experiment, \ie to provide an
accurate link between the large angular scales from \Cobe and the
first acoustic peak as measured by degree--scale experiments like {\sc
Boomerang}, {\sc Cbi}, {\sc Dasi}, {\sc Maxima}, {\sc Vsa}, has been
achieved.

\section{Cosmological constraints\label{cosmo}}
\begin{figure}[t]
\centering
\resizebox{\hsize}{!}
{
\includegraphics[clip]{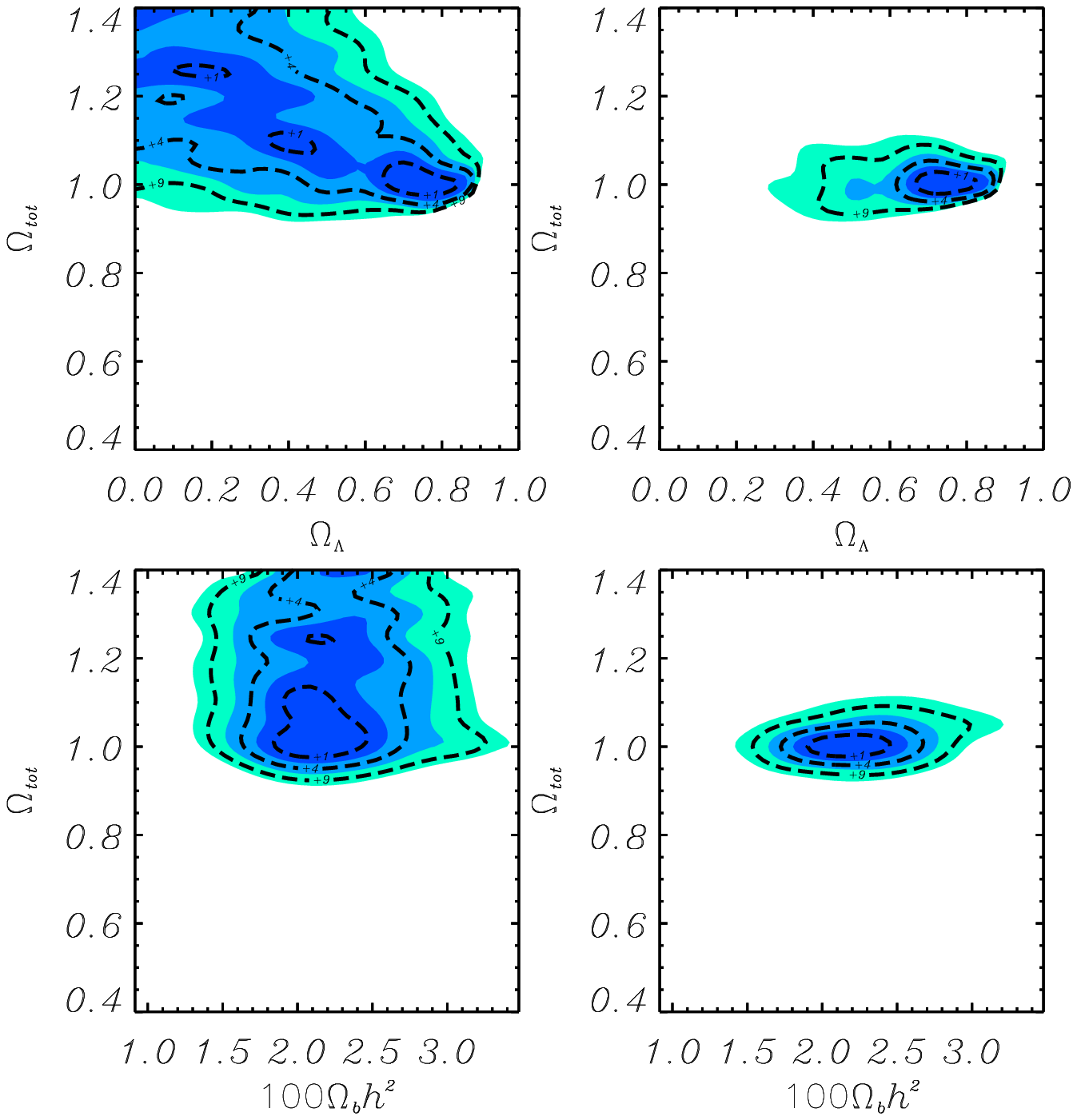} 
\includegraphics[clip]{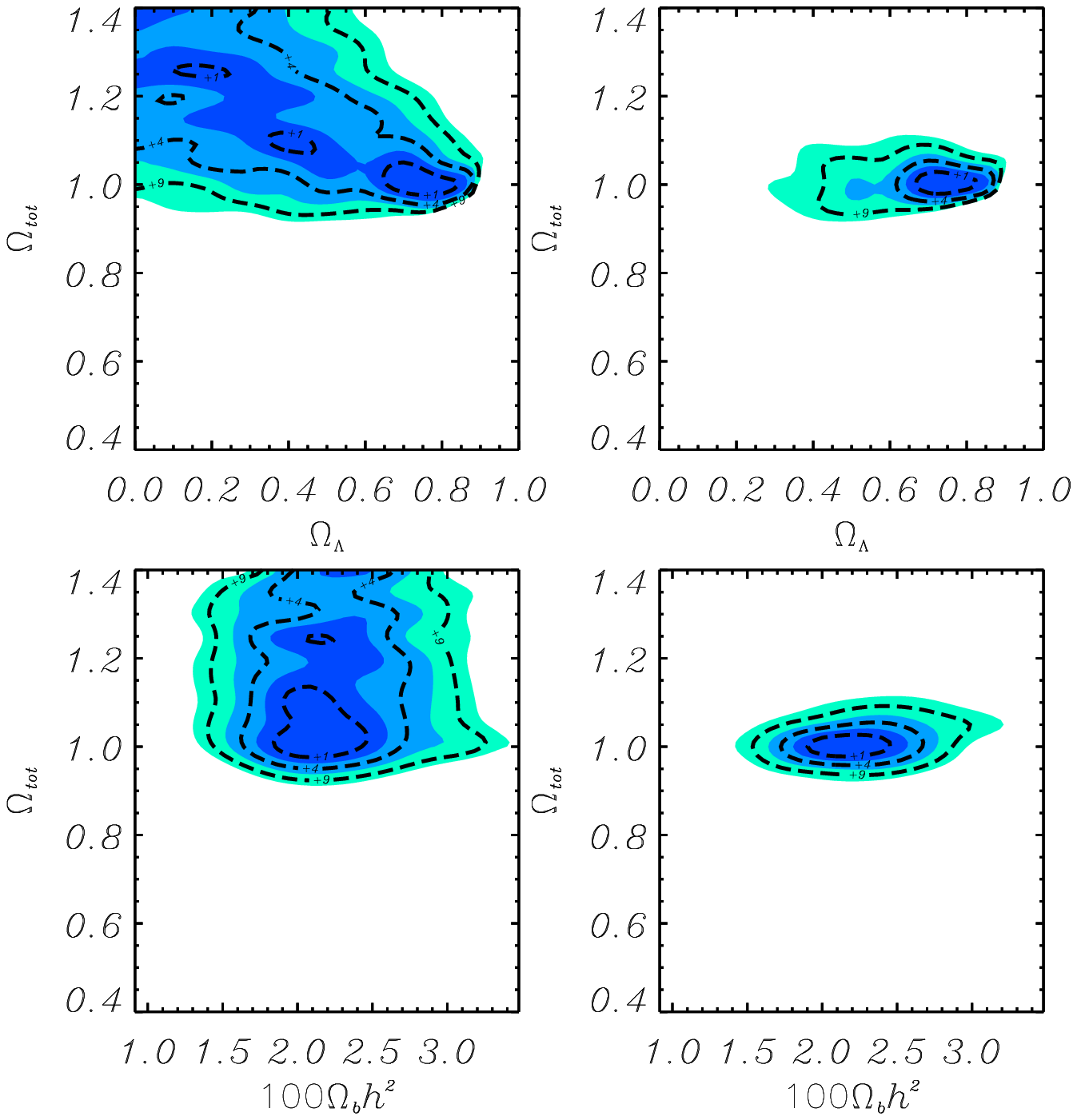} 
}
\caption{Likelihood contours between 3 of the cosmological parameters: 
baryonic density and cosmological constant in the abscissae and total
density as the ordinate. Greyscale corresponds to 2--D limits and
dashed line to 1--D contours with the Gaussian equivalent of 1, 2, and
$3\,\sigma$ thresholds. On the left, only the constraints from
\Archeops and other CMB experiments, identical to those in
Fig.~\ref{figure_cl}, are used. On the right panels, adding the prior
on the Hubble constant $H_0=72\pm8 \rm\,km/s/Mpc$ (68\% CL, Freedman
\etal 2001) reduces significantly the allowed values of the
cosmological parameters.}  \label{figure_param}
\end{figure}

Cosmological constraints can be placed on adiabatic cold dark matter
models with passive power-law initial fluctuations.  Because \Archeops
power spectrum has small bins in $\ell$ and large $\ell$ coverage down
to \Cobe scales, it provides a precise determination of the first
acoustic peak in terms of position at the multipole $l_{\rm
peak}=220\pm 6$, height and width.  Using a large grid of cosmological
adiabatic inflationary models described by 7 parameters, one can
compute their likelihood with respect to the datasets. An analysis of
Archeops data in combination with other CMB datasets constrains the
baryon content of the Universe to a value $\Ombhh =
0.022^{+0.003}_{-0.004}$ which is compatible with Big-Bang
nucleosynthesis (O'Meara \etal 2001)\cite{omeara} and with a similar
accuracy (Fig.~\ref{figure_param}).  Using the recent HST
determination of the Hubble constant~\cite{freedman} leads to tight
constraints on the total density, {\it e.g.}  $\OmT
=1.00^{+0.03}_{-0.02}$, \ie the Universe is flat.  An excellent
absolute calibration consistency is found between {\mbox{\sc{Cobe}}},
\Archeops and other CMB experiments (Fig.~\ref{figure_cl}). Finally,
an analysis adding data from other CMB experiments (\Cobe, \Boomerang,
\Dasi, \Maxima, \Vsa and \Cbi), the HST prior and an additionnal prior
on the value of $\sigma_8$ (see~\cite{douspis} for details) leads to
tight constraints on the quintessence parameters:
$\Omega_Q=0.70_{-0.17}^{+0.10}$ and $w_Q=-1^{+0.25}$ (95\% CL)
comforting the flat-$\Lambda$ cosmological model.

All these measurements are fully compatible with inflation--motivated
cosmological models. In particular, the best fit model shown in
Fig.~\ref{figure_cl} is close to the mean likelihood Universe
characterised by $\OmT=1.00^{+0.03}_{-0.02}$,
$\OmL=0.72^{+0.08}_{-0.06}$, $\Ombhh=0.021^{+0.001}_{-0.003}$,
$h=0.69^{+0.06}_{-0.06}$, $n=0.96^{+0.02}_{-0.04}$,
$Q=19.2\,\mathrm{\mu K}$, $\tau= 0$, obtained with \Archeops and other
CMB experiments and with the HST and $\tau=0$ priors.  Moreover, the
constraints shown in Fig.~\ref{figure_param}~(right), leading to a
value of $\OmL = 0.73^{+0.09}_{-0.07}$ for the dark energy content, is
independent from and in agreement with supernov{\ae} measurements
\cite{perlmutter} if a flat Universe is
assumed.

\section{Diffuse galactic plane emission maps\label{maps}}
\begin{figure}[h!]
\resizebox{\hsize}{!}
{
\includegraphics[clip,angle=90]{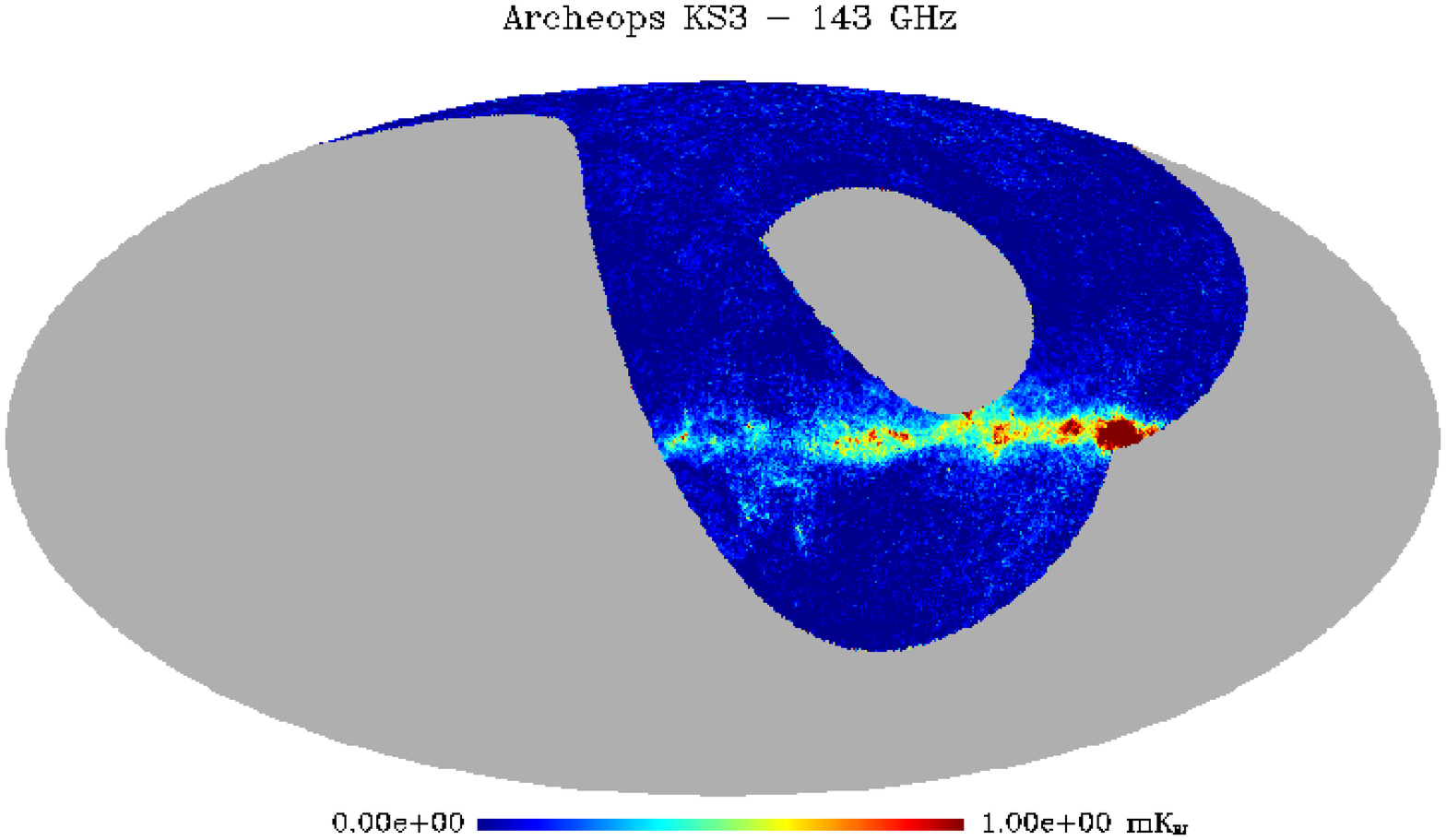}
\includegraphics[clip,angle=90]{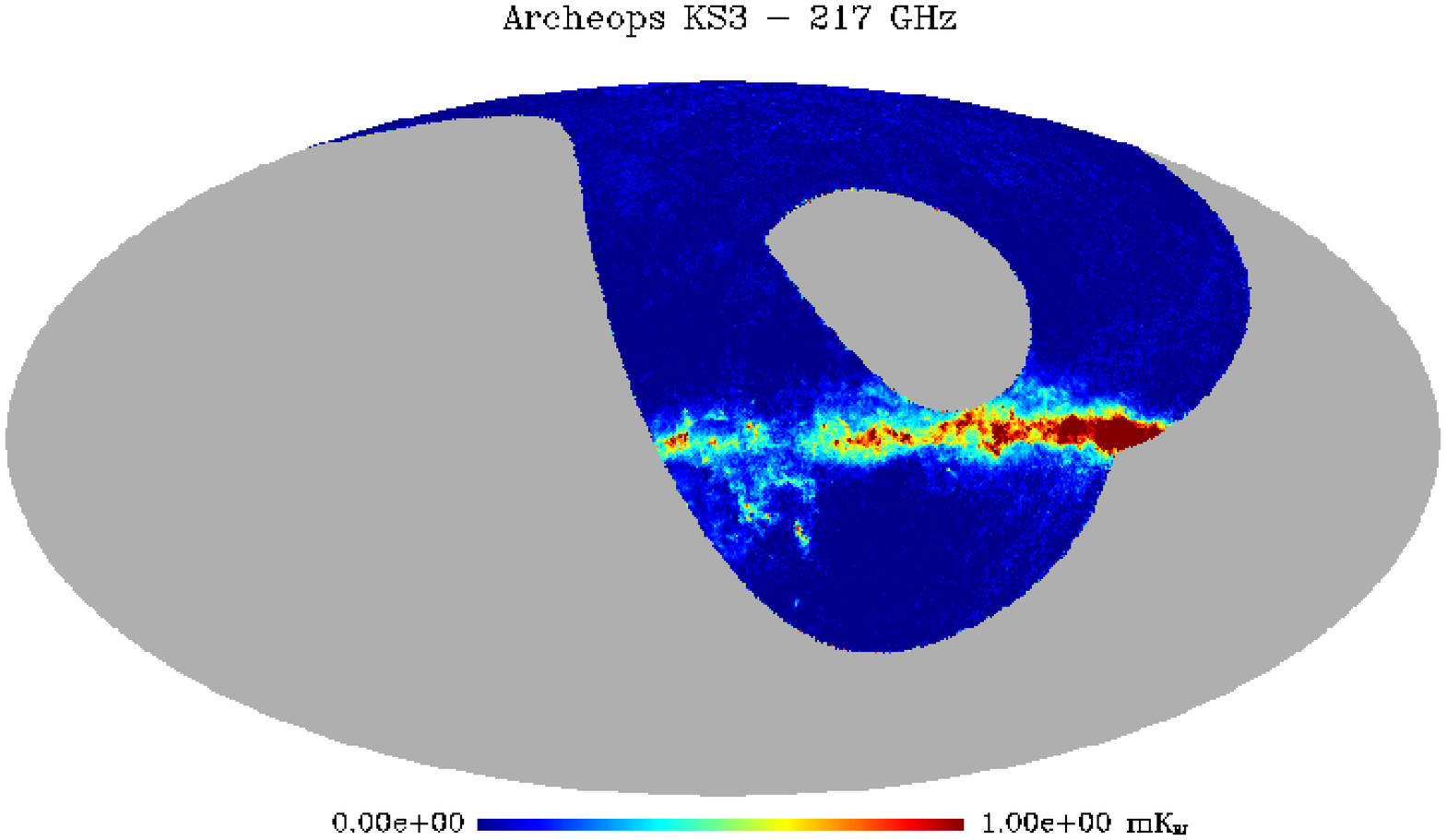}} \\
\resizebox{\hsize}{!}
{
\includegraphics[clip,angle=90]{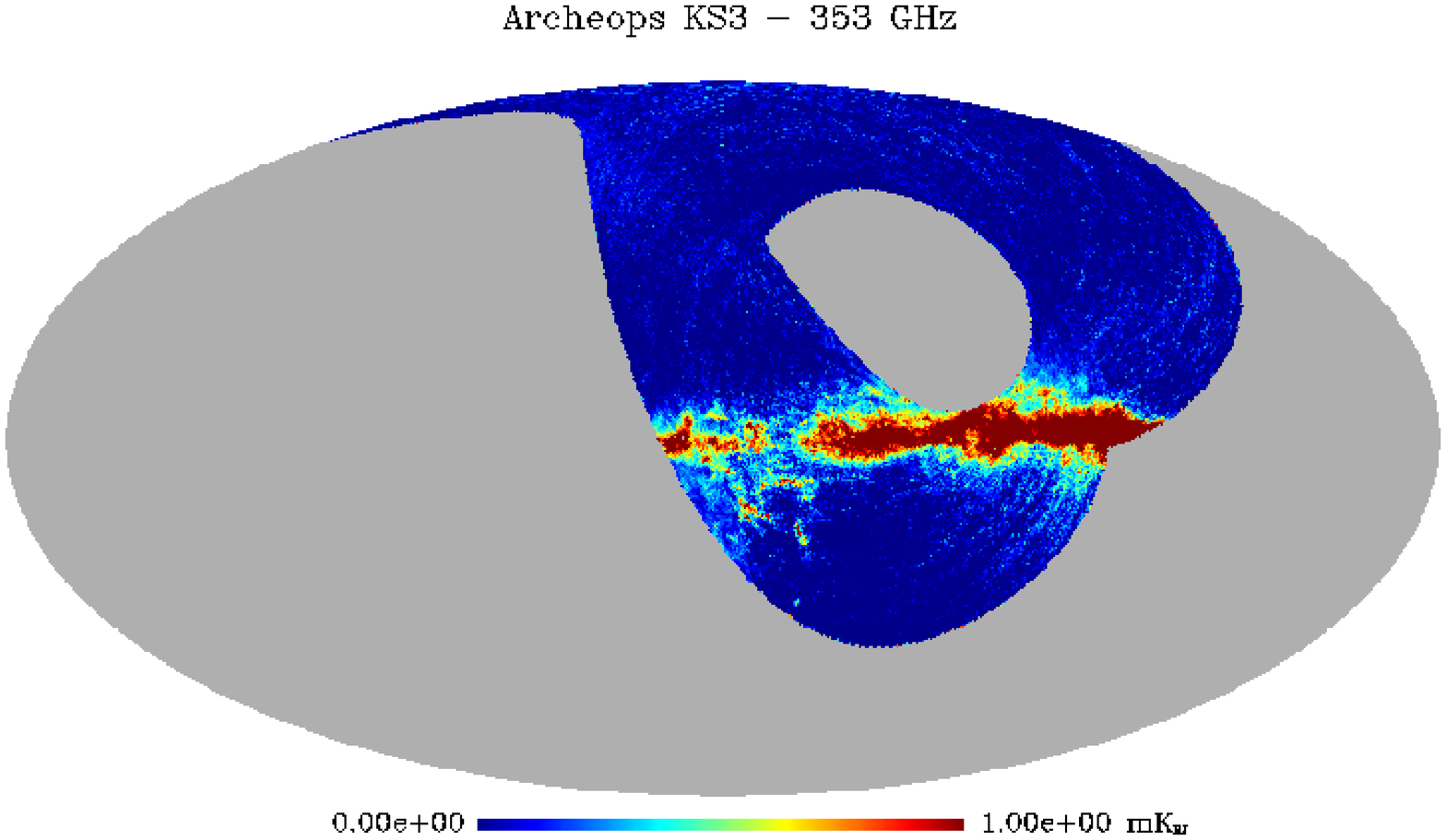}
\includegraphics[clip,angle=90]{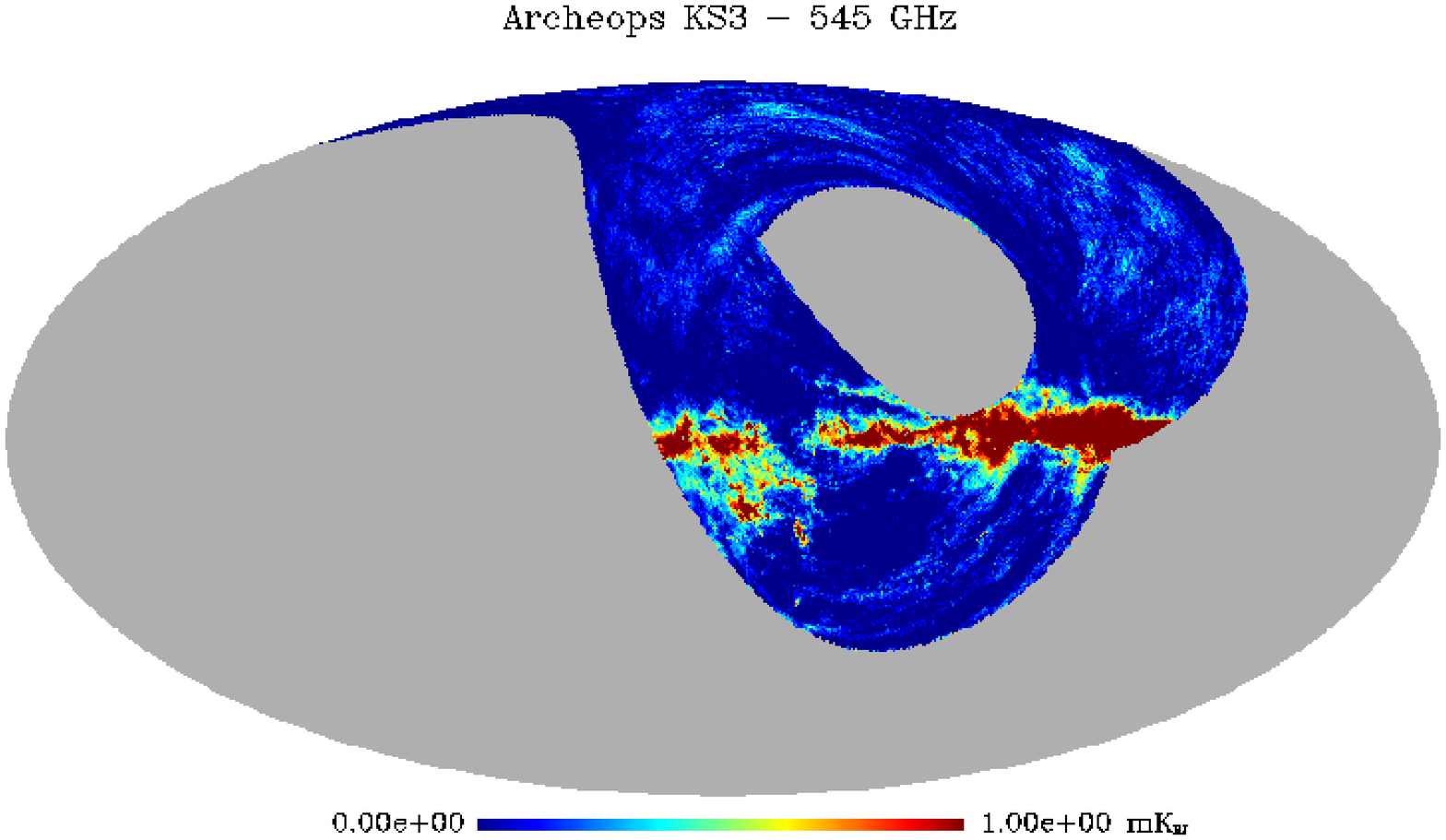}
}
\caption{Archeops Galactic maps at 143 GHz (top--left), 217 GHz (top--right), 
353 GHz (bottom--left) and 545 GHz (bottom--right). The four maps are
shown between 0 and 1 $\mathrm{mK_{RJ}}$.\label{fig_maps}}
\end{figure}

A specific data analysis pipeline has been designed to optimize the
maps in the Galactic plane region. We use a destriping method that
estimates the low frequency drifts by minimizing the cross-scan
variations in the map and introduces a 2$f_{spin}$ highpass
filter. The maps obtained by coadding all bolometers at each channel
are shown in Fig.~\ref{fig_maps}. These maps are the first Galactic
plane high resolution maps at these frequencies on such a wide area of
the sky. They show new dense bright regions, such as in the Taurus
complex, on the lower--left part and also diffuse emission at large
Galactic latitude.

\section{Dust polarisation detection\label{polar}}
The six 353 GHz Archeops channels are three Ortho Mode Transducers
(hereafter OMT)~\cite{boifot,chattopadhyay} composed by a pair of
bolometers having a common entry horn, the light is separated in its
two linear polarizations with a polarizer beam splitter, one
polarization is transmitted to the first bolometer and the second one
reflected to the second bolometer. The sum of two coupled bolometers
measures the total intensity while their difference measures the $Q$
Stokes parameters in the OMTs eigen basis. The three OMT are oriented
at 60 degrees from each other following the recommendations
of~\cite{couchot} to optimize the polarization reconstruction. Details
on the Archeops polarization experimental setup and data analysis will
be found in~\cite{benoit_polar}.

\begin{figure}[h!]
\resizebox{\hsize}{!}
{
\includegraphics[clip,angle=90]{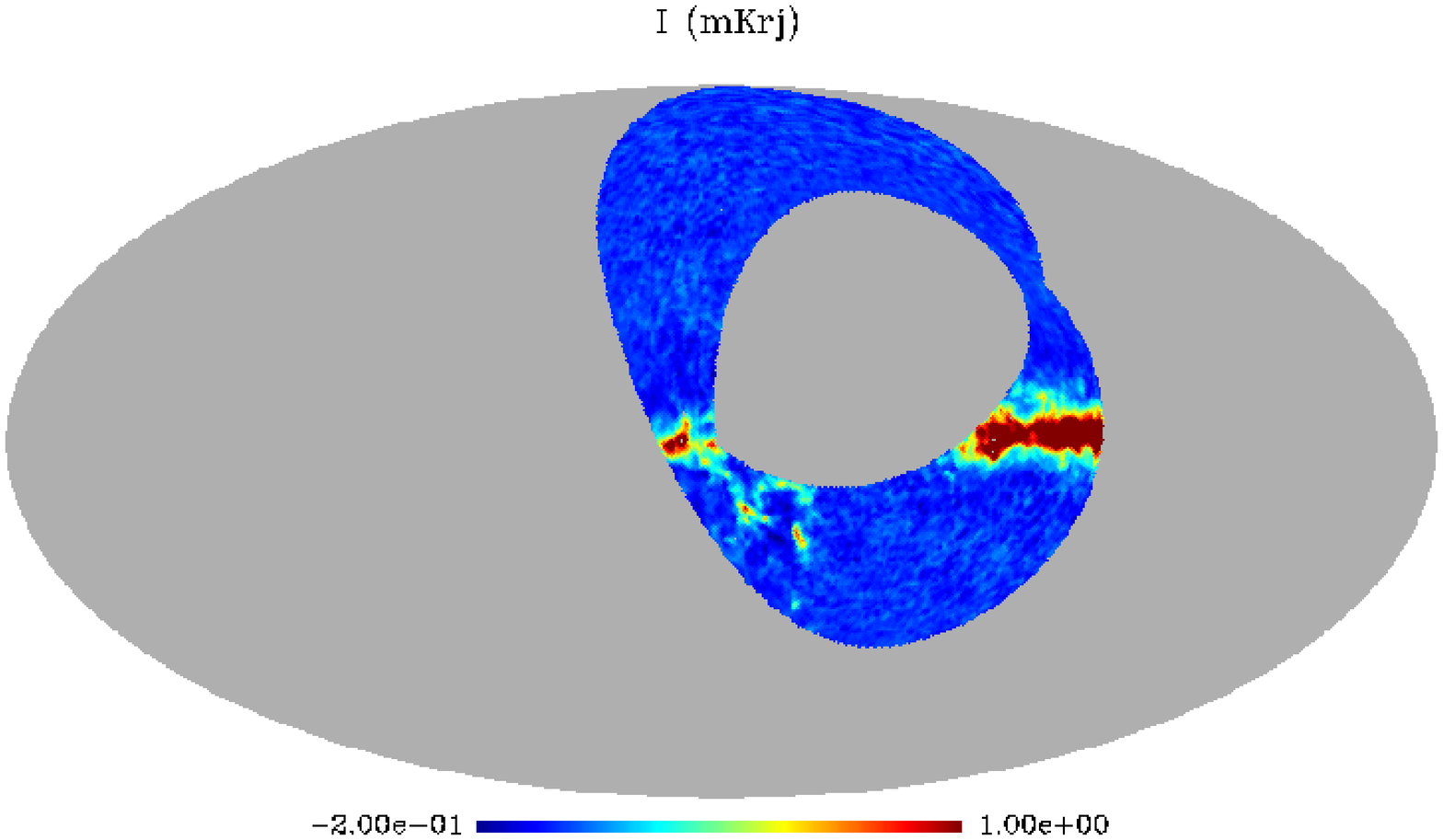}
\includegraphics[clip,angle=90]{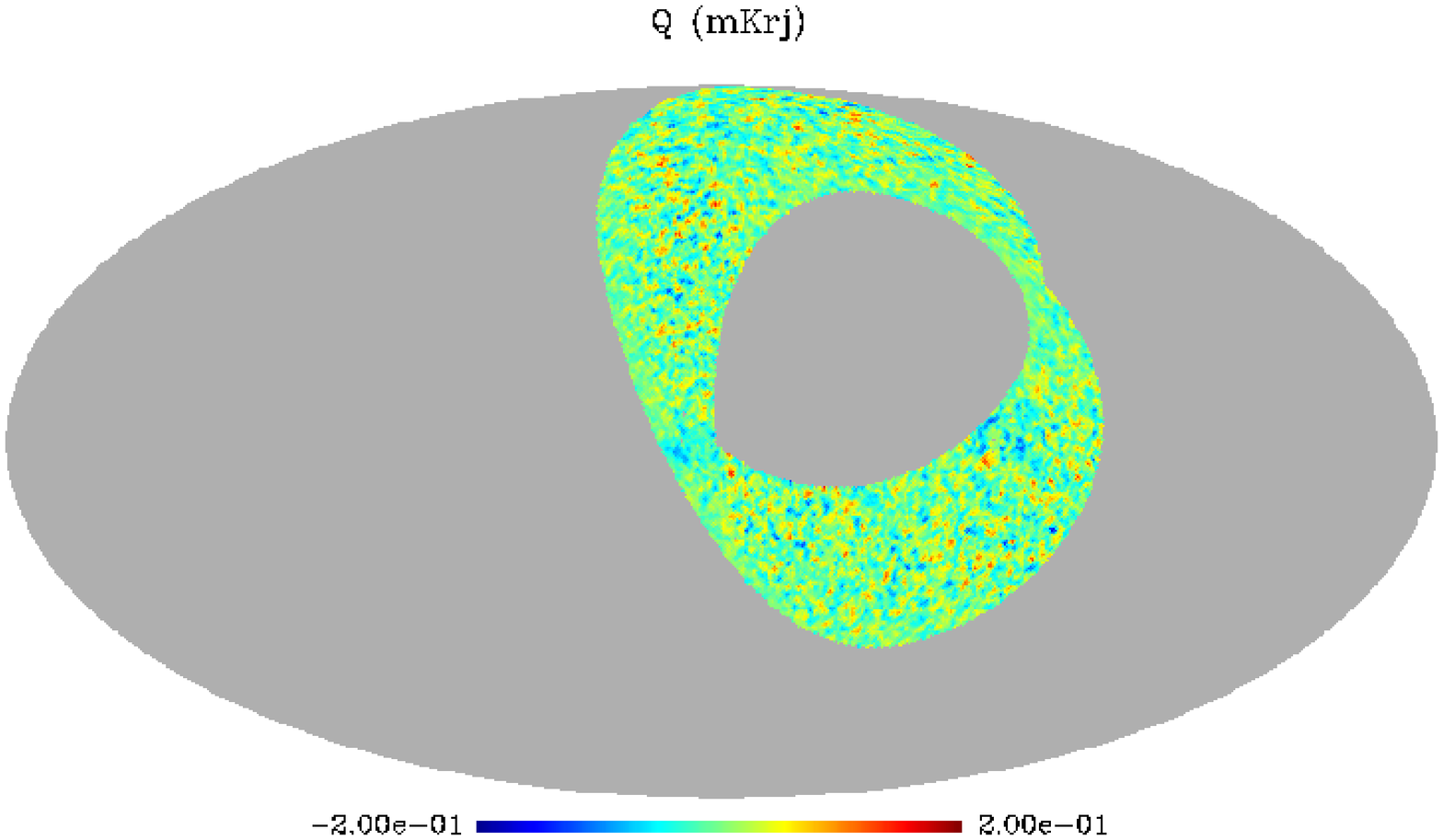}
\includegraphics[clip,angle=90]{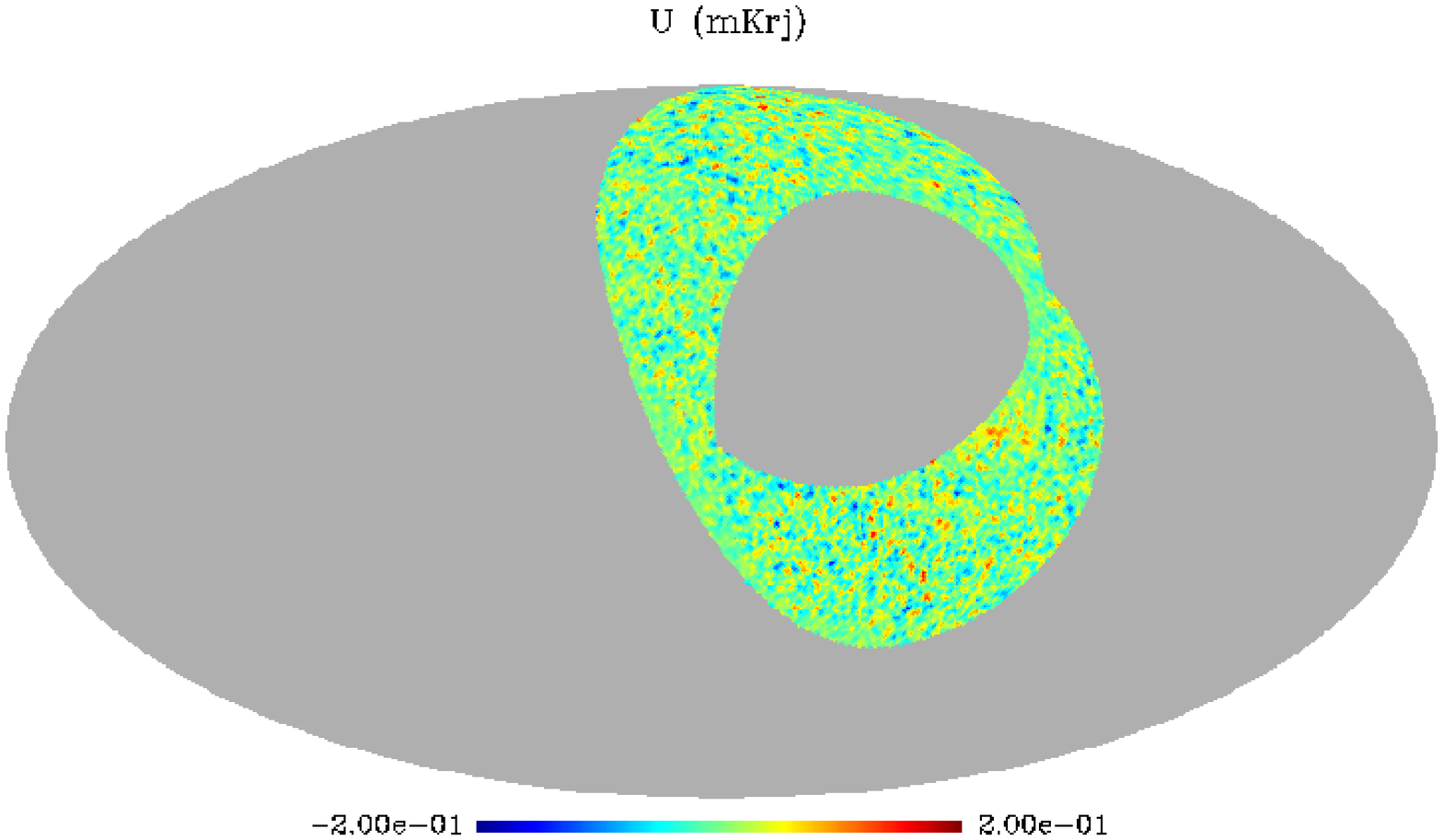}
\includegraphics[clip,angle=90]{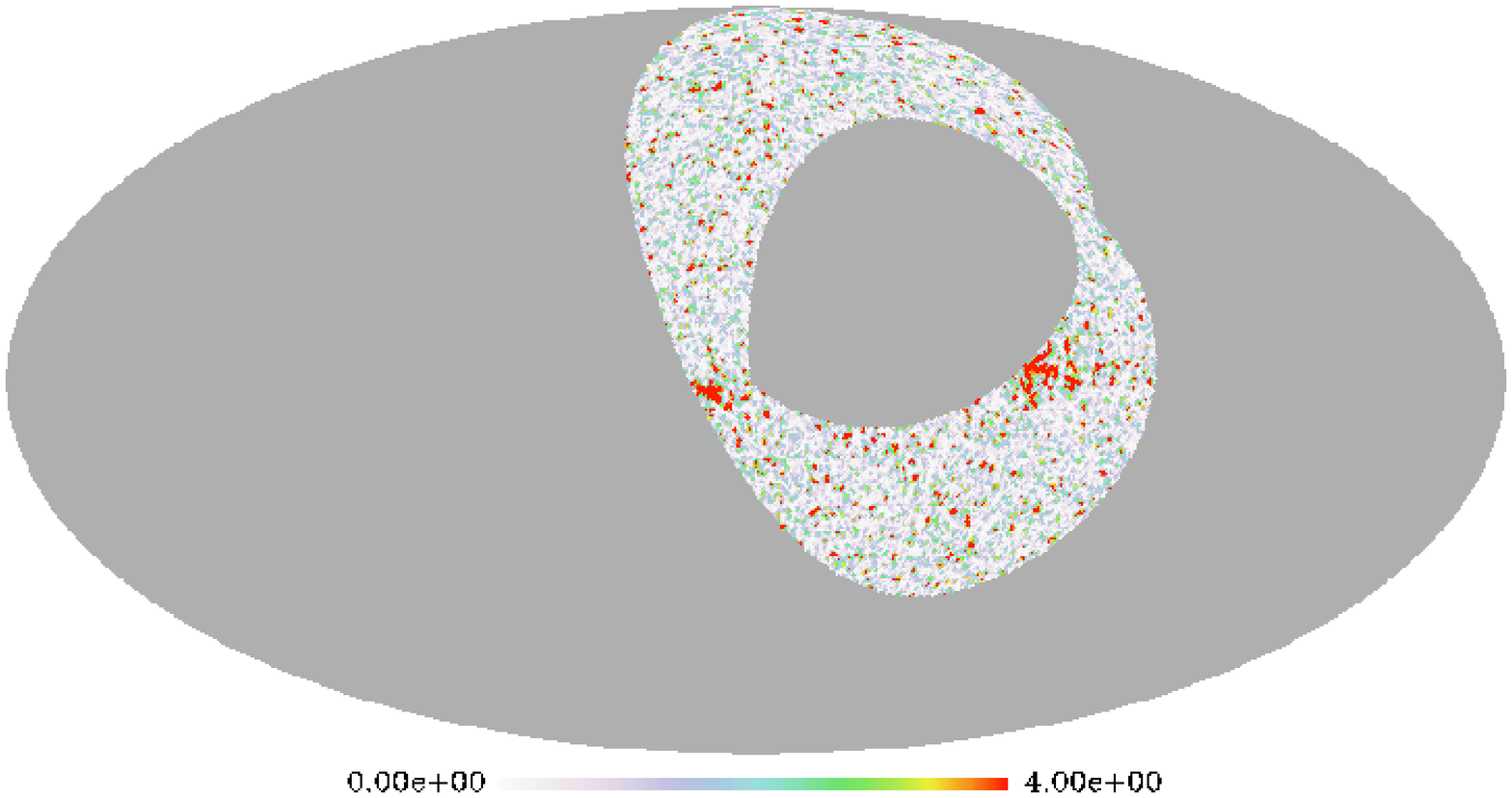}
}
\caption{Archeops $I$, $Q$, $U$ and normalized squared polarized intensity 
maps (from left to right). The color scales ranges are
$[-2,1]\mathrm{mK_{RJ}}$ for the $I$ map, $[-2,2]\mathrm{mK_{RJ}}$ for
the $Q$ and $U$ maps and $[0,4]$ for the normalized squared polarized
intensity map. In the latter map, red colour therefore corresponds to
a three $\sigma$ detection.\label{polmaps}}
\end{figure}

A specific analysis pipeline is used for the polarisation
analysis. Cross-calibration between two coupled channels is obtained
by intercomparing the large signal coming from Galactic latitude
profiles from different bolometers. Simulations show that the method
leads to an accuracy on the cross-calibration better than 1\% which is
sufficient to measure galactic dust polarization in an unbiased
way. The time stream data are filtered using a combination of the
method described in~\cite{amblard}, used for the \Archeops CMB
analysis and wavelet shrinkage techniques~\cite{juan}. This ensures
that no ringing is created in the Galactic plane region. The filtered
polarization channels are then combined to produce maps of $I$, $Q$
and $U$ with a pixel size of 27.5 arcmin. (HEALPIX nside=128). The
maps are clipped to remove pixels containing less than 100 time
samples corresponding to a $I$ noise level of
$143~\mu\mathrm{K_{RJ}}$. The maps are shown in Fig.~\ref{polmaps}
along with a map of the normalized squared polarized intensity $(Q^2 +
U^2)/(\sigma_Q^2 + \sigma_U^2)$. Twice this quantity is statistically
distributed like a $\chi^2$ with 2 degrees of freedom. The 68, 95.4,
99.7\% CL of the mapped quantity correspond to 1.1, 3.1, 5.8
respectively.

Seven polarized Galactic dense clouds are detected at the 3$\sigma$
level with sizes ranging from $2.3$ to $21.6$ deg$^2$ and polarization
degree ranging from $7.5\pm 1.7$\% to $23.2\pm 11$\%. The precise
characteristics of these clouds are given in~\cite{benoit_polar}. A
search for diffuse galactic polarization shows significant (3 to
4$\sigma$) coherent polarization levels of a few percents in the
Galactic plane, even after masking the dense clouds discussed
above. The detected clouds appear to belong to two complexes, one in
Cassiopeia that includes the CasA supernova remnant (the point source
was however not projected in the maps), the other in the southern part
of Gem OB1. The polarization orientation
is found to be coherent between clouds and diffuse regions and nearly
orthogonal to the Galactic plane. These results are in agreement with
expectations from starlight polarization measurements~\cite{fosalba},
the presence of dense polarized regions provides evidence for a
powerful grain alignement mechanism thoughout the interstellar
medium~\cite{hilde_95} and a coherent magnetic field coplanar to the
Galactic plane and following the spiral arms. Interestingly, the observed 
part of the Cygnus complex is not found to be significantly polarized. 
Projection effects along the line of sight of sight may explain the lack 
of detectable polarization from Cygnus region.
Extrapolating these results to high Galactic latitude
indicates that interstellar dust polarized emission is the major
foreground for PLANCK-HFI CMB polarization measurements as suggested
by~\cite{prunet}.

\section*{Conclusions}

Constraints on various cosmological parameters (Benoît \etal
2003b)\cite{benoit_params} have been derived by using the \Archeops
data alone and in combination with other measurements.  The measured
power spectrum (Benoît \etal 2003a)\cite{benoit_cl} matches the \Cobe
data and provides for the first time a direct link between the
Sachs--Wolfe plateau and the first acoustic peak, because of the large
sky coverage that greatly reduces the sample variance.  The measured
spectrum is in good agreement with that predicted by simple inflation
models of scale--free adiabatic peturbations and a flat--$\Lambda$
Universe assumption. Finally let us note that these results were
obtained with only half a day worth of data.  Precise maps of the
Galactic dust emission in the Galactic plane region have been obtained
and show new sources and diffuse emission.
\Archeops also provides the first measurement of galactic dust polarization 
at 353 GHz showing dense polarized clouds and diffuse
polarization. The coherence of the polarization direction suggests the
presence of a coherent magnetic field coplanar to the Galactic plane.

Use of all available bolometers and of a larger sky fraction should
yield an even more accurate and broader CMB power spectrum in the near
future. The large experience gained on this balloon--borne experiment
is providing a large feedback to the \Planck -- \Hfi data processing
community.

\Acknowledgements{We pay tribute to Pierre Faucon, the manager of the CNES 
launching team in Kiruna, who was a key figure for the success of the 
experiment. We wish to thank ASI, CNES, and PNC for their continued 
support. Healpix package~\cite{healpix} was used
throughout this work.}

%
\end{document}